\documentclass[a4paper]{amsart}

\usepackage{graphics,amssymb,enumerate}
\usepackage{hyperref}
\usepackage[dvips]{graphicx}
\usepackage[all]{xy}

\def\bd{\begin{displaymath}}
\def\ed{\end{displaymath}}
\def\be{\begin{equation}}
\def\ee{\end{equation}}

\def\cal{\mathcal}
\newtheorem{thm}{Theorem}
\newtheorem{prop}{Proposition}

\newtheorem{cor}{Corollary}

\newtheorem{rem}{Remark}

\theoremstyle{definition}

\theoremstyle{remark}

\Large
\begin{document}

\title{ Darboux polynomials for Lotka-Volterra systems in three dimensions}

\author{Yiannis T. Christodoulides and   Pantelis A.~Damianou}

\address{Department of Mathematics and Statistics\\
University of Cyprus\\
P.O.~Box 20537, 1678 Nicosia\\Cyprus}

 \email{ychris@ucy.ac.cy,  damianou@ucy.ac.cy}

\Large
\begin{abstract}

We consider Lotka-Volterra systems in three dimensions depending on three real
parameters. By using elementary algebraic methods we classify the Darboux polynomials
(also known as second integrals) for such systems for various values of the parameters,
and give the explicit form of the
corresponding cofactors. More precisely, we show that a Darboux
polynomial of degree greater than one is reducible. In fact, it is
a product of linear Darboux polynomials and first integrals.

\end{abstract}

\date{16 October 2008}
\maketitle

\Large

\section{Introduction}

The Lotka-Volterra model is a basic model of predator-prey
interactions. The model was developed independently by Alfred
Lotka (1925), and Vito Volterra (1926).  It forms the basis for
many models used today in the analysis of population dynamics. It
has other applications in Physics, e.g. laser Physics, plasma
Physics (as an approximation to the Vlasov-Poisson equation), and
neural networks. In three dimensions it describes the dynamics of
a biological system where three species interact.

 The
most general form of Lotka-Volterra equations is
$$\dot x_i = \varepsilon_i x_i + \sum_{j=1}^n a_{ij} x_i x_j, \ \
i=1,2, \dots , n.$$

 We consider  Lotka-Volterra equations without linear terms ($\varepsilon_i=0$), and where
the matrix of interaction coefficients  $A= ( a_{ij} ) $ is  skew-symmetric. This is a natural assumption
related to the principle that crowding inhibits growth.

The most famous special case of Lotka-Volterra system is the  KM
system (also known as the Volterra system) defined by

\begin{equation}
\dot x_i = x_i(x_{i+1}-x_{i-1}) \qquad i=1,2, \dots,n, \label{a1}
\end{equation}
where $x_0 \!= x_{n+1} \!=0$.
 It was first solved by Kac and
van-Moerbeke in \cite{km}, using a discrete version of inverse
scattering due to Flaschka \cite{flaschka}. In \cite{moser} Moser
gave a solution of the system using the method of continued
fractions, and in the process he constructed action-angle
coordinates. Equations (\ref{a1}) can be considered as a
finite-dimensional approximation of the Korteweg-de Vries (KdV)
equation.  The variables $x_i$ are an intermediate step in the
construction of the action-angle variables for the Liouville model
on the lattice. This system has a close connection with the Toda
lattice,

\begin{eqnarray*}
\dot{a}_i&=&a_i (b_{i+1}-b_{i}) \quad  \ \ \ \  i=1,\ldots,n-1 \\
\dot{b}_i&=&2(a_i^2-a_{i-1}^2)  \quad \ \ \ \ i=1,\ldots,n.
\end{eqnarray*}

In fact, a transformation of H\'{e}non connects the two systems:
\begin{eqnarray*}
a_i&=&-\frac{1}{2}\sqrt{x_{2i}x_{2i-1}} \quad \ \ \ \ i=1,\ldots,n-1 \\
b_i&=&\frac{1}{2}(x_{2i-1}+x_{2i-2}) \quad  \ \ \ \ i=1,\ldots,n.
\end{eqnarray*}

The systems which we consider are all integrable in the sense of
Liouville.  In other words, there are enough integrals in
involution to ensure the complete integrability of the system.

Any constant value of a first integral defines a submanifold
which is invariant under the flow of the Hamiltonian vector field.
A second integral is a function which is constant on a specific
level set.  While a first integral satisfies $\dot{f}=0$, a second
integral is characterized by  the property  $\dot{f}=\lambda f$, for  some function
$\lambda$ which is called the cofactor of $f$.  Second integrals
are also called special functions, stationary solutions, and in
the case of polynomials, eigenpolynomials, or, more frequently,
Darboux polynomials. In systems which have a Lie theoretic origin
(e.g. the full Kostant Toda lattice), they arise from
semi-invariants of group actions. The importance of Darboux
polynomials lies in the following simple fact. If $f$ and $g$ are
relatively prime Darboux polynomials, with the same cofactor, then
their quotient is a first integral.  We propose to understand the
behavior of a system based on the algebraic properties of its
Darboux polynomials.

As a starting point we  consider the system

\be  \label{rst}
\begin{array}{lcl}
\dot x_1&=&x_1(rx_2+sx_3)\cr
 \dot x_2&=&x_2(-rx_1+tx_3)\cr
 \dot x_3&=&x_3(-sx_1-tx_2)
\end{array}
 \ee
 where $r$, $s$, $t$ $\in\Bbb R$.

Our main result is the following:

\begin{thm}\label{thm0}
An arbitrary Darboux polynomial of the system (\ref{rst}) is
reducible. In fact, it is a product of linear Darboux polynomials.
\end{thm}

The method of proof that we use follows  the  approach of  Labrunie in \cite{labrunie} for the so called
ABC system.

The system (\ref{rst}) is Hamiltonian. We define the following
quadratic Poisson bracket in $\Bbb R^3$ by the formula

\be \label{poisson} \pi = r x_1 x_2 \frac {\partial }{\partial
x_1} \wedge \frac {\partial} {\partial x_2}+ s x_1 x_3 \frac
{\partial }{\partial x_1} \wedge  \frac {\partial} {\partial x_3}+
t x_2 x_3 \frac {\partial }{\partial x_2} \wedge  \frac {\partial}
{\partial x_3} \ . \ee

Generically, the rank of this Poisson bracket is 2 and it
possesses a Casimir given by $F=x_1^t x_2^{-s}x_3^r$.  The
function $H=x_1+x_2+x_3$ is always a constant of motion. In fact,
taking $H$ as the Hamiltonian and using the Poisson bracket
(\ref{poisson}) we obtain equations (\ref{rst}).

Lotka-Volterra systems have been studied extensively, see e.g.   \cite{bermejo},
\cite{grammaticos}, \cite{plank}. The Darboux
method of finding integrals of finite dimensional vector fields
and especially for various types of Lotka-Volterra systems has
been used by several authors, e.g. \cite{cf}, \cite{cl},  \cite{labrunie}
\cite{mp}, \cite{mo1}, \cite{mo2}.

The paper is organized as follows. In Section 2, we recall a few
basic facts about Darboux polynomials. In Section 3 we prove
Theorem \ref{thm0} under general conditions for $r$, $s$, $t$, and
we also give the explicit form of the cofactors.  Section 4 deals
with the case $s=t$.  We did not examine other such cases since
the method of proof is identical with these two cases.  Finally in
Section 5 we present in detail three examples which include the
 open and periodic KM-system in three dimensions.

\section{Darboux polynomial preliminaries} Consider a system of
ordinary differential
equations\begin{equation}\label{2}\frac{dx_i}{dt}=v_i(x_1(t),...,
x_n(t)),\hspace{6mm}i=1,...,n,\end{equation}where the functions
$v_i$ are smooth on a domain $U\subset \Bbb K^n$. Here $\Bbb
K=\Bbb R$ or $\Bbb K=\Bbb C$, and we denote by $\Bbb K[\bf x]$,
${\bf x}=(x_1,...,x_n)$, the ring of polynomials in $n$ variables
over $\Bbb K$. Let $\phi :I\rightarrow U$ be a solution of
(\ref{2}) defined on an open non-empty interval $I$ of the real
axis. A continuous function $F:U\rightarrow\Bbb R$ is called a
{\it first integral} of system (\ref{2}) if it is constant along
its solution, i.e. if the function $F\circ\phi$ is constant on its
domain of definition for arbitrary solution $\phi$ of (\ref{2}).
When $F$ is differentiable, it is a first integral of system
(\ref{2}) if\begin{equation}\label{3}L_{\bf v}(F)=\sum
_{i=1}^{n}v_i({\bf x})\frac {\partial F}{\partial x_i}({\bf
x})=0,\end{equation}where $L_{\bf v}$ is the {\it Lie derivative}
along the vector field ${\bf v}=(v_1,...,v_n)$. If $A$ is any
function of ${\bf x}$, then the Lie derivative of $A$ is the {\it
time derivative} of $A$, i.e. $\dot A=\frac {dA}{dt}=L_{\bf
v}(A)$. The vector field generates a {\it flow} $\phi _t$ that
maps a subset $U$ of $\Bbb K^n$ to $\Bbb K^n$ in such a way that a
point in $U$ follows the solution of the differential equation.
That is, $\dot\phi ({\bf x})(t)={\bf v}(\phi ({\bf x}))$ $\forall$
${\bf x}\in U$. The time derivative is also called the {\it
derivative along the flow} since it describes the variation of a
function of $\bf x$ with respect to $t$ as $\bf x$ evolves
according to the differential system.\vspace{1mm}

Many first integral search techniques, such as the Prelle-Singer
procedure, are based on the Darboux polynomials. A polynomial
$f\in\Bbb K[{\bf x}]$ is called a {\it Darboux polynomial} of
system (\ref{2}) if
\begin{equation}\label{4}L_{\bf v}(f)=\lambda f,\end{equation}for
some polynomial $\lambda\in\Bbb K [{\bf x}]$, which is called the
cofactor of $f$. When $\lambda =0$, the Darboux polynomial is a
first integral; $f$ is said to be a proper Darboux polynomial if
$\lambda \not= 0$. Let $f_1$, $f_2$ be Darboux polynomials with
cofactors $\lambda _1$, $\lambda _2$, respectively. It is easy to
verify that:\\(i) The product $f_1f_2$ is also a Darboux
polynomial, with cofactor $\lambda _1+\lambda _2$, and\\(ii) If
$\lambda _1=\lambda _2=\lambda$ then the sum $f_1+f_2$ is also a
Darboux polynomial, with cofactor $\lambda $.\\ The following
propositions (\cite{goriely}) give some more elementary but
important properties of Darboux polynomials.

\begin{prop}\label{prop1} Let $f$,
$g\in\Bbb K [{\bf x}]$ be non-zero and coprime (i.e. they do not
have common divisors different from constants). Then, $f\backslash
g$ is a rational first integral if and only if $f$ and $g$ are
Darboux polynomials with the same cofactor $\lambda\in\Bbb K [{\bf
x}]$.
\end{prop}

\begin{prop}\label{prop2} $(i)$ All irreducible
factors of a Darboux polynomial are Darboux polynomials,\\$(ii)$
Suppose that the system $(\ref{2})$ is homogeneous of degree $m$,
i.e. all $v_i$ are homogeneous of degree $m$, and let $f$ be an
arbitrary Darboux polynomial of $(\ref{2})$ with cofactor
$\lambda$. Then $\lambda$ is homogeneous of degree $m-1$, and all
homogeneous components of $f$ are Darboux polynomials of
$(\ref{2})$ with cofactor $\lambda$.
\end{prop}

Thus, the search for Darboux polynomials can be restricted to
irreducible polynomials, and, if the system is homogeneous, to
homogeneous polynomials. Since the dynamical system (\ref{rst}) is
homogeneous of degree 2, the cofactor of any Darboux polynomial of
the system will be a linear combination of the variables $x_1$,
$x_2$, $x_3$. It follows that any Darboux polynomial $f$ of system
(\ref{rst}) will
satisfy\begin{equation}\label{5}L(f)=x_1(rx_2+sx_3)\frac {\partial
f}{\partial x_1}+x_2(-rx_1+tx_3)\frac {\partial f}{\partial
x_2}+x_3(-sx_1-tx_2)\frac {\partial f}{\partial x_3}=(\alpha
x_1+\beta x_2+\gamma x_3)f,\end{equation}where $\alpha$, $\beta$,
$\gamma$ are constants. If it is not clear from the context, we
shall denote these constants by $\alpha (f)$, $\beta (f)$, $\gamma
(f)$ respectively.

\section {Darboux polynomials of the Lotka-Volterra system }
We carry out our analysis aiming at maximum generality, that is,
imposing as few conditions on the parameters $r$, $s$, $t$ as
possible. In this section we make such assumptions in propositions
\ref{prop9}, \ref{prop10} and Theorem \ref{thm1}, however, as we
note in remark \ref{rem3}, one can obtain the results by making
assumptions about the cofactor of the Darboux polynomial instead
of the parameters. An important role in this work plays the
homogeneity property, as can be seen in the following two
propositions.

\begin{prop}\label{prop3} Let $f$ be a homogeneous Darboux
polynomial of system $(\ref{rst})$ of degree $m$. If $\gamma
(f)\not= 0$, then $f$ has no $x_3^m$ term so that
$f(x_1,x_2,x_3)=x_1\phi (x_1,x_2,x_3)+x_2\psi (x_1,x_2,x_3)$.
\end{prop}
\hspace{-6mm}Proof. Since the polynomial $f$ is homogeneous, we
use Euler's identity\begin{equation}\label{6}x_1\frac {\partial
f}{\partial x_1}+x_2\frac {\partial f}{\partial x_2}+x_3\frac
{\partial f}{\partial x_3}=mf.\end{equation}Using equation
(\ref{6}) we substitute for $x_1\frac {\partial f}{\partial x_1}$
in equation (\ref{5}) to obtain\[x_2(-rx_1+tx_3-rx_2-sx_3)\frac
{\partial f}{\partial x_2}+x_3(-sx_1-tx_2-rx_2-sx_3)\frac
{\partial f}{\partial x_3}=(\alpha x_1+(\beta -mr)x_2+(\gamma
-ms)x_3)f.\]Setting $x_1=0$, $x_2=0$, and letting
$F(x_3)=f(0,0,x_3)$ we
have\begin{equation}\label{7}-sx_3^2F^{\prime}(x_3)=(\gamma
-ms)x_3F(x_3).\end{equation}If $s=0$, $\gamma\not= 0$, equation
(\ref{7}) implies that $F=0$. Otherwise, if $s\not= 0$ we have
$F(x_3)=\kappa x_3^{m-\gamma /s}$, for some constant $\kappa$.
Since $f$ is homogeneous of degree $m$, the only term containing
only $x_3$ is necessarily $x_3^m$. Thus, if $\gamma\not= 0$ we
must have $F=0$ also in this case, and the proposition is
proved.\hfill$\Box$\vspace{1mm}

 We shall use the following notation: for a polynomial $f=f(x_1,x_2,x_3)$
 we denote $\hat f=f|_{x_1=0}$, $\bar f=f|_{x_2=0}$, $\breve f=f|_{x_3=0}$.
 We denote $N_m=\{1,2,...,m\}$, ${\cal N}_m=N_m\cup\{
0\}$, and for any number $r$, $N_mr=\{nr:n\in N_m\}$.

\begin{prop}\label{prop4} Let $f$ be a homogeneous Darboux
polynomial of degree $m$. If $\gamma (f)\not= 0$, then
$s=0\Rightarrow x_2|f$, $t=0\Rightarrow x_1|f$, $s\not= 0$ and
$\gamma \notin N_ms \Rightarrow x_2|f$, $t\not= 0$ and $\gamma
\notin N_mt\Rightarrow x_1|f$.\\ We also have the following
statements for $\alpha$ and $\beta$:\\  If $\beta (f)\not= 0$,
then $r=0\Rightarrow x_3|f$, $t=0\Rightarrow x_1|f$, $r\not= 0$
and $\beta\notin N_mr\Rightarrow x_3|f$, $t\not= 0$ and
$\beta\notin -N_mt\Rightarrow x_1|f$.\\If $\alpha (f)\not= 0$,
then $r=0\Rightarrow x_3|f$, $s=0\Rightarrow x_2|f$, $r\not= 0$
and $\alpha\notin -N_mr\Rightarrow x_3| f$, $s\not= 0$ and
$\alpha\not= -N_ms\Rightarrow x_2|f$.
\end{prop}
\hspace{-6mm}Proof. We prove the statements for $\gamma$. The
proof of the statements for $\alpha$ and $\beta$ is similar.
Assume that $\gamma \not= 0$. Then, it follows from Proposition
\ref{prop3} that $f=x_1\phi _1+x_2\psi _1$, where $\phi _1=\phi
_1(x_1,x_2,x_3)$, $\psi _1=\psi _1 (x_1,x_2,x_3)$ are either
homogeneous polynomials of degree $m-1$, or zero (but they are not
both zero). Setting this in equation (\ref{5})
yields\begin{equation}\label{8}x_1L(\phi _1)+x_2L(\psi _1)=(\alpha
x_1+\beta x_2+\gamma x_3-rx_2-sx_3)x_1\phi _1+(\alpha x_1+\beta
x_2+\gamma x_3+rx_1-tx_3)x_2\psi _1.\end{equation}Setting $x_2=0$
in equation (\ref{8}) we have\[x_1\overline {L(\phi _1)}=(\alpha
x_1+(\gamma -s)x_3)x_1\bar\phi _1.\]The operator $\phi
_1\rightarrow\bar\phi _1$ commutes with the derivations with
respect to $x_1$ and $x_3$, and therefore we
obtain\begin{equation}\label{9}sx_1x_3\Big (\frac {\partial
\bar\phi _1}{\partial x_1}-\frac {\partial \bar\phi _1}{\partial
x_3}\Big )=(\alpha x_1+(\gamma -s)x_3)\bar\phi _1
.\end{equation}If $s=0$ then $\bar\phi _1 =0$, which implies that
$\phi _1$ is divisible by $x_2$ and that $f=x_1\phi _1+x_2\psi _1$
is divisible by $x_2$. Suppose now that $s\not= 0$, $\deg \bar\phi
_1=\deg \phi _1=m-1$, and that $\gamma\not= ns$, $n\in N_m$. The
r.h.s. of (\ref{9}) is divisible by $x_1$, and since $\gamma
-s\not= 0$, it follows that $x_1|\bar\phi _1$. Let $\bar\phi _1
=x_1\phi _2$, where $\phi _2$ is a homogeneous polynomial of
degree $m-2$. Then, we have\[\frac {\partial\bar\phi _1}{\partial
x_1}=x_1\frac {\partial\phi _2}{\partial x_1}+\phi
_2,\hspace{8mm}\frac {\partial\bar\phi _1}{\partial x_3}=x_1\frac
{\partial \phi _2}{\partial x_3},\]and from (\ref{9}) we
obtain\[sx_1x_3\Big (\frac {\partial\phi _2}{\partial x_1}-\frac
{\partial \phi _2}{\partial x_3}\Big )=(\alpha x_1+(\gamma
-2s)x_3)\phi _2.\]Since $\gamma -2s\not= 0$, $\phi _2$ is
divisible by $x_1$. Continuing in the same way we obtain
\[sx_1x_3\Big (\frac {\partial\phi _{m-1}}{\partial x_1}-\frac
{\partial \phi _{m-1}}{\partial x_3}\Big )=(\alpha x_1+(\gamma
-(m-1)s)x_3)\phi _{m-1},\]where $\deg \phi _{m-1}=1$, and
$x_1|\phi _{m-1}$. Thus, $\phi _{m-1}=const.\hspace{0.5mm}x_1$,
and from the above equation we have $sx_3=\alpha x_1+(\gamma
-(m-1)s)x_3$. By equating coefficients we obtain $\gamma =ms$,
which is a contradiction. Therefore, we must have $\bar\phi _1=0$,
which implies that $f$ is divisible by $x_2$.

Setting $x_1=0$ in (\ref{8}) and using (\ref{5}) we
obtain\[tx_2x_3\Big (\frac {\partial \widehat\psi _1}{\partial
x_2}-\frac {\partial\widehat\psi _1}{\partial x_3}\Big ) =(\beta
x_2+(\gamma -t)x_3)\widehat\psi _1.\]If $t=0$ then $\widehat\psi
_1=0$, hence $\psi _1$ is divisible by $x_1$ and so $f$ is
divisible by $x_1$. Suppose that $t\not= 0$, $\deg \widehat\psi
_1=\deg\psi _1=m-1$, and $\gamma\not= nt$, $n\in N_m$. Then it can
be shown in a similar way as above that $\psi _1$ is divisible by
$x_1$, which implies that $f$ is divisible by $x_1$, and the
proposition is proved.\hfill$\Box$\vspace{2mm}

This leads to the characterization of the cofactors of Darboux
polynomials of system (\ref{rst}), as follows.

\begin{prop}\label{prop5} Let $f$ be a homogeneous Darboux
polynomial of degree $m$. We have either $\gamma (f)=0$, or
$\gamma (f)=\gamma _1s$, $\gamma _1\in N_m$, or $\gamma (f)=\gamma
_2t$, $\gamma _2\in N_m$, or $\gamma (f)=\gamma _1s+\gamma _2t$,
$\gamma _2\in\{ 1,2,...,m-1\}$, $\gamma _1\in N_{m-\gamma _2}$.
\end{prop}
\hspace{-6mm}Proof. Since $f$ is a Darboux polynomial it satisfies
$L(f)=(\alpha x_1+\beta x_2+\gamma x_3)f$. Suppose that
$\gamma\not= 0$ and $\gamma \not= ns$, $n\in N_m$. Then by
proposition \ref{prop4} $f$ is divisible by $x_2$, that is
$f=x_2f_1$ for some homogeneous polynomial $f_1$ of degree $m-1$
and we have\[L(f_1)=((\alpha +r)x_1+\beta x_2+(\gamma
-t)x_3)f_1.\]Suppose that $\gamma (f_1)\not= 0$, i.e. $\gamma\not=
t$, and that $\gamma (f_1)\not=ns$, $n\in N_{m-1}$, that is
$\gamma\not= ns+t$, $n\in N_{m-1}$. Then, again by proposition
\ref{prop4} it follows that $f_1$ is divisible by $x_2$, and
writing $f_1=x_2f_2$ we obtain\[L(f_2)=((\alpha +2r)x_1+\beta
x_2+(\gamma -2t)x_3)f_2.\]If $\gamma \not= 2t$, and $\gamma\not=
ns+2t$, $n\in N_{m-2}$, then $f_2$ is divisible by $x_2$.
Continuing in the same way, after $m-1$ steps we
obtain\begin{equation}\label{10}L(f_{m-1})=((\alpha
+(m-1)r)x_1+\beta x_2+(\gamma
-(m-1)t)x_3)f_{m-1},\end{equation}where $\deg f_{m-1}=1$. If
$\gamma\not= (m-1)t$ and $\gamma \not= s+(m-1)t$, then
$x_2|f_{m-1}$, and thus $f_{m-1}=const.\hspace{1mm}x_2$. From
equation (\ref{10}) we then have $-rx_1+tx_3=(\alpha
+(m-1)r)x_1+\beta x_2+(\gamma -(m-1)t)x_3$, and by equating
coefficients we obtain $\gamma =mt$. We therefore conclude that we
have either $\gamma =0$, or $\gamma =ns$, or $\gamma =nt$, $n\in
N_m$, or $\gamma =\gamma _1s+\gamma _2t$, $\gamma _2=1,2,...,m-1$,
$\gamma _1\in N_{m-\gamma _2}$, and the proposition is
proved.\hfill$\Box$

\begin{rem}\label{rem1} We note that in the proof of proposition
$\ref{prop5}$
we can make the successive assumptions $\gamma (f)\not= nt$ $(n\in
N_m$), $\gamma (f_1)\not= nt$, $(n\in N_{m-1}$),..., $\gamma
(f_{m-1})\not= nt$, $(n\in N_1$), which imply that the respective
functions are divisible by $x_1$. We obtain the same result also
in this case, in particular the relation $\gamma _1s+\gamma _2t$
with the conditions $\gamma _1=1,2,...,m-1$, $\gamma _2\in
N_{m-\gamma _1}$, which are the same with the conditions stated in
the proposition.
\end{rem}

\begin{prop}\label{prop6}Let $f$ be a homogeneous
Darboux polynomial of degree $m$. We have:\\ $(a)$ $\alpha (f)=0$,
or $\alpha (f)=-\alpha _1r$, $\alpha _1\in N_m$, or $\alpha
(f)=-\alpha _2s$, $\alpha _2\in N_m$, or $\alpha (f)=-\alpha
_1r-\alpha _2s$, $\alpha _2=1,2,...,m-1$, $\alpha _1\in
N_{m-\alpha _2}$.\\ (b) $\beta (f)=0$, or $\beta (f)=\beta _1r$,
$\beta _1\in N_m$, or $\beta (f)=-\beta _2t$, $\beta _2\in N_m$,
or $\beta (f)=\beta _1r-\beta _2t$, $\beta _2=1,2,...,m-1$, $\beta
_1\in N_{m-\beta _2}$.
\end{prop}

\hspace{-6mm}Proof. The proof is similar to the proof of
proposition \ref{prop5}.\vspace{2mm}

The following propositions give further analysis on the cofactors,
and their relation with the parameters and the form of the Darboux
polynomials.

\begin{prop}\label{prop7} Let $r$, $s$, $t$ be non-zero,
$r\backslash s=q_1$, $r\backslash t=q_2$, and $s\backslash t=q_3$.
Let $f$ be a homogeneous Darboux polynomial of degree $m$, and
$\alpha _1$, $\alpha _2$, $\beta _1$, $\beta _2$, $\gamma _1$,
$\gamma _2$ the integers which appear in propositions
$\ref{prop5}$ and $\ref{prop6}$.\\
$(a)$ If $\alpha _1+(\alpha _2-j)\frac 1{q_1}\notin {\cal
N}_{m-j}$ and $\beta _1-(\beta _2-j)\frac 1{q_2}\notin {\cal
N}_{m-j}$, for $j=0,1,2,...,m-1$,
then $\alpha _2=\beta _2$.\\
(b) If $(\alpha _1-j)q_1+\alpha _2\notin {\cal N}_{m-j}$ and
$\gamma _1+(\gamma _2-j)\frac 1{q_3}\notin {\cal N}_{m-j}$, for
$j=0,1,2,...,m-1$,
then $\alpha _1=\gamma _2$.\\
(c) If $-(\beta _1-j)q_2+\beta _2\notin {\cal N}_{m-j}$ and
$(\gamma _1-j)q_3+\gamma _2\notin {\cal N}_{m-j}$, for
$j=0,1,2,...,m-1$, then $\beta _1=\gamma _1$.
\end{prop}
\hspace{-6mm}Proof. We prove statement $(a)$. The proof of
statements (b) and (c) is similar. If $\alpha _2$ or $\beta _2$ is
non-zero, then by hypothesis we have $\alpha (f)=-(\alpha
_1+\alpha _2\frac 1{q_1})r\not= 0$ and $\alpha (f)\not= -nr$,
$n\in N_m$, or $\beta (f)=(\beta _1-\beta _2\frac 1{q_2})r\not= 0$
and $\beta (f)\not= nr$, $n\in N_m$, respectively. In either case,
it follows from proposition \ref{prop4} that $f$ is divisible by
$x_3$. We can write $f=x_3f_1$, for some homogeneous polynomial
$f_1$ of degree $m-1$, and we have\[L(f_1)=((\alpha +s)x_1+(\beta
+t)x_2+\gamma x_3)f_1\]\[=((-\alpha _1r-(\alpha _2-1)s)x_1+(\beta
_1r-(\beta _2-1)t)x_2+\gamma x_3)f_1.\]By the same argument as
above, if we do not have $\alpha _2(f_1)=\beta _2(f_1)=0$, i.e. if
we do not have $\alpha _2=\beta _2=1$, then we have either $\alpha
(f_1)=-(\alpha _1+(\alpha _2-1)\frac 1{q_1})r\not= 0$ and $\alpha
(f_1)\not= -nr$, $n\in N_{m-1}$, or $\beta (f_1)=(\beta _1-(\beta
_2-1)\frac 1{q_2})r\not= 0$ and $\beta (f_1)\not= nr$, $n\in
N_{m-1}$, and $f_1$ is divisible by $x_3$. Continuing in the same
way, after $m-1$ steps we obtain\begin{equation}\label{11}L
(f_{m-1})=((-\alpha _1r-(\alpha _2-(m-1))s)x_1+(\beta _1r-(\beta
_2-(m-1))t)x_2+\gamma x_3)f_{m-1},\end{equation}where $\deg
f_{m-1}=1$. If we do not have $\alpha _2=\beta _2=m-1$, then it
follows by our assumptions that $x_3|f_{m-1}$, which implies that
$f_{m-1}=const.\hspace{1mm}x_3$ and $f=const.\hspace{1mm}x_3^m$.
However, $\alpha _2(x_3)=\beta _2(x_3)=1$, and by simple
properties of Darboux polynomials it follows that $\alpha
_2(f)=\beta _2(f)=m$. Therefore, we must have $\alpha _2=\beta
_2=n$, for some integer $n\in\{ 0,1,2,...,m\}$, and the
proposition is proved.\hfill$\Box$

\begin{prop}\label{prop8} Let $r$, $s$, $t$
be non-zero, $r\backslash s=q_1$, and $r\backslash t=q_2$. Let $f$
be a proper Darboux polynomial, homogeneous of degree $m$, with
$\gamma (f)=0$, and let $\alpha _1$, $\alpha _2$, $\beta _1$,
$\beta _2$ be the integers which appear in proposition
$\ref{prop6}$.\\
(a) If $\alpha _1\not= 0$ and $\alpha _1q_1+\alpha _2\notin {\cal
N}_m$, or $\beta _1\not= 0$ and $-(\beta _1q_2-\beta _2)\notin
{\cal N}_m$,
then $s=-pt$, for some positive rational number $p$.\\
(b) If $\alpha _1=\beta _1=0$, $(\alpha _2-j)\frac 1{q_1}\notin
N_{m-j}$ and $-(\beta _2-j)\frac 1{q_2}\notin N_{m-j}$, for
$j=0,1,2,...,m-1$, then we have $f=x_3^{\alpha _2}I$ where $I$ is
a first integral.
\end{prop}
\hspace{-6mm}Proof. $(a)$ Suppose $\alpha _1\not= 0$ and $\alpha
_1q_1+\alpha _2\notin {\cal N}_m$. The other case is similar. Then
we have $\alpha (f)=-(\alpha _1q_1+\alpha _2)s\not= 0$ and $\alpha
(f)\not= -ns,$ $n\in N_m$. From proposition \ref{prop4} it follows
that $f$ is divisible by $x_2$, so that $f=x_2f_1$ for some
homogeneous polynomial $f_1$ of degree $m-1$, and we have
\begin{equation}\label{12}L(f_1)=((\alpha +r)x_1+\beta
x_2-tx_3)f_1.\end{equation}Equation (\ref{12}) shows that $f_1$ is
a Darboux polynomial with $\gamma (f_1)=-t$. However, from
proposition \ref{prop5} we have $\gamma (f_1)=\gamma _1s+\gamma
_2t$ for non-negative integers $\gamma _1$, $\gamma
_2\in\{0,1,2,...,m-1\}$. Therefore, $\gamma _1s+\gamma _2t=-t$,
which is possible only if $\gamma _1\not= 0$, in which case
$s=-\frac {(1+\gamma _2)}{\gamma _1}t$, and the statement is
proved with $p=\frac {1+\gamma _2}{\gamma _1}$.\\
(b) Suppose that $\alpha _1=\beta _1=0$. Since $f$ is a proper
Darboux polynomial with $\gamma (f)=0$ we must have $\alpha
_2\not= 0$ or $\beta _2\not= 0$, and our assumptions imply that in
fact $\alpha _2=\beta _2$ (see proposition \ref{prop7}). We have
$\alpha (f)=-\alpha _2\frac 1{q_1}r\not= 0$ and $\alpha (f)\not=
-nr$, $n\in N_m$. It follows from proposition \ref{prop4} that $f$
is divisible by $x_3$. So $f=x_3f_1^{\prime}$ for some polynomial
$f_1^{\prime}$ of degree $m-1$, and we have\[L
(f_1^{\prime})=(-(\alpha _2-1)sx_1-(\beta
_2-1)tx_2)f_1^{\prime}.\]By the same argument, if $\alpha
_2-1=\beta _2-1\not= 0$, then $f_1^{\prime}$ is divisible by
$x_3$. Continuing in the same way, we find that $f=x_3^{\alpha
_2}I$ for some first integral $I$ ($I\equiv 1$ if $\alpha _2=\beta
_2=m$), and the proposition is proved.\hfill$\Box$\vspace{2mm}

These results allow us to characterize the Darboux polynomials of
system (\ref{rst}).

\begin{prop}\label{prop9}Let $f$ be a Darboux polynomial
of system $(\ref{rst})$, homogeneous of degree $m$. If $s=0$ then
\begin{equation}\label{13}f=x_2^{\gamma _2}f_1,\end{equation}
where $f_1$ is a Darboux polynomial with $\gamma (f_1)=0$. If $s$,
$t$ are non-zero and $N_m s\cap N_m t=\emptyset$, then we
have\begin{equation}\label{14}f=x_1^{\gamma _1}x_2^{\gamma
_2}f_2,\end{equation}where $f_2$ is a Darboux polynomial with
$\gamma (f_2)=0$. Here, the non-negative integers $\gamma _1$,
$\gamma _2$ are such that $\gamma (f)=\gamma _1s+\gamma _2t$.
\end{prop}
\hspace{-6mm}Proof. If $\gamma (f)=0$, then the result in each
case follows by setting $\gamma _1 =\gamma _2=0$, $f_1=f_2=f$.
Suppose that $\gamma (f)\not= 0$ and $s=0$. Then, by proposition
\ref{prop4} $f$ is divisible by $x_2$, and writing
$f=x_2f_1^{\prime}$ we have \[L(f_1^{\prime})=((\alpha
+r)x_1+\beta x_2+(\gamma -t)x_3)f_1^{\prime}.\] Let this procedure
be repeated as many times as it can, and let $\gamma _2$ be the
number of times that it can. We have $f=x_2^{\gamma _2}f_1$, where
$f_1$ is a Darboux polynomial with $\gamma (f_1)=\gamma -\gamma
_2t=0$ since we had to stop the division procedure by $x_2$, and
equation (\ref{13}) is proved. Suppose now that $\gamma (f)\not=
0$, $s$, $t$ are non-zero and $N_m s\cap N_m t=\emptyset$. Thus
$\gamma\notin N_m s$ or $\gamma\notin N_m t$. Let us consider the
case $\gamma \notin N_m s$. The case $\gamma\notin N_m t$ is
similar. Then $f$ is divisible by $x_2$ and as before we have
$f=x_2^{\gamma _2}f_2^{\prime}$, where $f_2^{\prime}$ is a Darboux
polynomial with $\gamma (f_2^{\prime})=\gamma -\gamma _2t$. Since
we had to stop the division procedure by $x_2$, we must have
either $\gamma (f_2^{\prime})=0$, in which case equation
(\ref{14}) is satisfied with $\gamma _1=0$ and $f_2=f_2^{\prime}$,
or $\gamma (f_2^{\prime})=\gamma _1s$, for some $\gamma _1\in
N_m$. In the latter case $\gamma (f_2^{\prime})\notin N_m t$ and
$f_2^{\prime}$ is divisible $\gamma _1$ times by $x_1$, that is,
$f_2^{\prime}=x_1^{\gamma _1}f_2$, $\gamma (f_2)=0$, and equation
(\ref{14}) follows.\hfill$\Box$

\begin{rem}\label{rem2} The condition $N_m s\cap N_m
t= \emptyset$ implies that there do not exist integers $n_1$,
$n_2\in N_m$ such that $s=\frac {n_2}{n_1}t$. This condition is
satisfied in each of the following cases:\\ $(a)$ one of $s$, $t$
is positive and the other is negative,\\ (b) $s$, $t$ have the
same sign but one is rational and the other irrational,\\ (c) $s$,
$t$ have the same sign, they are both irrational, and their ratio
is irrational,\\ (d) $s$, $t$ have the same sign, they are both
rational, and $s/t<1/m$ or $s/t>m$,\\ (e) $s$, $t$ have the same
sign, they are both irrational, their ratio is rational, and
$s/t<1/m$ or $s/t>m$.
\end{rem}

\begin{rem}\label{rem3} In proposition $\ref{prop9}$, instead of the
condition $N_m s\cap N_m t=\emptyset$, we can make an alternative
assumption as follows. First let $s\backslash t=q_3$, and let
$f_k$, $k=0,1,2,...$, $f=f_0$, be a sequence of Darboux
polynomials as we describe below. We denote $\gamma (f_k)=\gamma
_1(f_k)s+\gamma _2(f_k)t$, $\gamma _1=\gamma _1(f)$, $\gamma
_2=\gamma _2(f)$. For $k=0,1,2,...,\gamma _1+\gamma _2-1$, we
suppose that
\begin{equation}\label{15} (i)\hspace{3mm}\gamma
_1(f_k)+\gamma _2(f_k)\frac 1{q_3}\notin {\cal
N}_{m-k}\hspace{8mm}or\hspace{8mm}(ii)\hspace{3mm}\gamma
_1(f_k)q_3+\gamma _2(f_k)\notin {\cal N}_{m-k}.\end{equation}In
particular, if $\gamma _1(f_k)=0$ then we require condition $(i)$
to hold, whereas if $\gamma _2(f_k)=0$ then we require condition
$(ii)$ to hold $($if $\gamma _1(f_k)\not= 0$ and $\gamma
_2(f_k)\not= 0$ then we can have either condition $(i)$ or
$(ii))$. If condition $(i)$ holds, then $\gamma (f_k)\not= 0$ and
$\gamma (f_k)\not= ns$, $n\in N_{m-k}$, which implies that $f_k$
is divisible by $x_2$. Thus $f_k=x_2f_{k+1}$, and $\gamma
_1(f_{k+1})=\gamma _1(f_k)$, $\gamma _2(f_{k+1})=\gamma
_2(f_k)-1$. If condition $(ii)$ holds, then $\gamma (f_k)\not= 0$
and $\gamma (f_k)\not= nt$, $n\in N_{m-k}$, which implies that
$x_1|f_k$. In this case we have $f_k=x_1f_{k+1}$, $\gamma
_1(f_{k+1})=\gamma _1(f_k)-1$, $\gamma _2(f_{k+1})=\gamma
_2(f_k)$. Following this procedure, after $\gamma _1+\gamma _2$
steps we obtain $f=x_1^{\gamma _1}x_2^{\gamma _2}f^{\prime}$,
where $\gamma (f^{\prime})=0$.
\end{rem}

The following proposition states similar results in terms of the
constants $\alpha$ and $\beta$. The proof is similar to the proof
of proposition \ref{prop9}.

\begin{prop}\label{prop10}Let $f$ be a homogeneous Darboux polynomial
of degree $m$, and let $\alpha _1$, $\alpha _2$, $\beta _1$,
$\beta _2$ be the integers which appear in proposition
$\ref{prop6}$.
\\ (i) If $s=0$ then $f=x_2^{\alpha _1}f_1$, where $f_1$ is a
Darboux polynomial with $\alpha (f_1)=0$.\\(ii) If $r$, $s$ are
non-zero and $-N_m r\cap (-N_m s)=\emptyset$ then $f=x_2^{\alpha
_1}x_3^{\alpha _2}f_2$, where $f_2$ is a Darboux polynomial with
$\alpha (f_2)=0$.\\(iii) If $r$, $t$ are non-zero and $N_m r\cap
(-N_m t)=\emptyset$ then $f=x_1^{\beta _1}x_3^{\beta _2}f_3$,
where $f_3$ is a Darboux polynomial with $\beta (f_3)=0$.
\end{prop}

We are now ready to prove the main result of this section.

\begin{thm}\label{thm1}Let $f$ be a Darboux polynomial
of system $(\ref{rst})$, homogeneous of degree $m$. Suppose that
either: (i) $s=0$ and $N_m r\cap (-N_m t)=\emptyset$, or (ii) $r$,
$s$, $t$ are non-zero, $N_m r\cap (-N_m t)=\emptyset $, $N_m s\cap
N_m t=\emptyset$, and $(-N_m r)\cap (-N_m s)=\emptyset$. (In
particular, condition (ii) is satisfied, for example, when $r>0$,
$t>0$ and $s<0$, or $r<0$, $t<0$ and $s>0$). Then, there exist
three non-negative integers $i$, $j$, $k$ and a polynomial first
integral I -which may be trivial- such
that\begin{equation}\label{16}f=x_1^ix_2^jx_3^kI\end{equation}and
\begin{equation}\label{17}
\alpha (f)=-rj-sk,\hspace{7mm}\beta (f)=ri-tk,\hspace{7mm}\gamma
(f)=si+tj.\end{equation}
\end{thm}
\hspace{-6mm}Proof. Consider the case $s=0$ and $N_m r\cap (-N_m
t)=\emptyset$. The other case is similar. We use equation
(\ref{13}) of proposition \ref{prop9} and the equations in
statements (i) and (iii) of proposition \ref{prop10} in the
following algorithm.
\\(1) Set $n=0$ and $f_n=f$.\\(2) Applying proposition
\ref{prop10} for $\alpha$ (statement (i)) yields\[f_n=x_2^{\alpha
_1}f_{n+1},\hspace{7mm}\alpha (f_{n+1})=0.\]If $f_{n+1}$ is a
first integral, go to the final step, else increment $n$ by
one.\\(3) Applying proposition \ref{prop10} for $\beta$ (statement
(iii)) yields\[f_n=x_1^{\beta _1}x_3^{\beta
_2}f_{n+1},\hspace{7mm}\beta (f_{n+1})=0.\]If $f_{n+1}$ is a
first integral, go to the final step, else increment $n$ by one.\\
(4)Applying proposition \ref{prop9} for $\gamma$ (eq. (\ref{13}))
yields\[f_n=x_2^{\gamma _2}f_{n+1},\hspace{7mm}\gamma
(f_{n+1})=0.\]If $f_{n+1}$ is a first integral, go to the final
step, else increment $n$ by one and return to step 2.\\(5) (Final
step) Set $I=f_{n+1}$ and using the sequence of equations linking
$f_l$ to $f_{l+1}$, $l=1,...,n$ given by the algorithm determine
the exponents $i$, $j$, $k$ in eq. (\ref{16}).

At every step one has $\deg f_{l+1}\leq $ $\deg f_l$; when three
consecutive terms of the sequence are of the same degree, they are
equal and $\alpha (f_l)=\beta (f_l)=\gamma (f_l)=0$, so $f_l$ is a
first integral. Thus the algorithm converges in a finite number of
steps. Equation (\ref{17}) follows from simple properties of
Darboux polynomials.

If condition (ii) holds, then the proof is the same but now in
steps 2 and 4 of the algorithm we use the equation in statement
(ii) of proposition \ref{prop10}, and equation (\ref{14}) of
proposition \ref{prop9}, respectively.\hfill$\Box$

\section {The case $s=t$} In this section we study the case $s=t$,
which is not covered by Theorem \ref{thm1} in the previous
section. It can be seen that in this case $x_1+x_2$ is an
additional linear Darboux polynomial of system (\ref{rst}), with
 cofactor $sx_3$. Therefore, polynomials of the form
$f=x_1^ix_2^jx_3^k(x_1+x_2)^l$, where $i$, $j$, $k$, $l$ are
non-negative integers, are Darboux polynomials. We show that a
Darboux polynomial will have this form with $l>0$, provided its
cofactor satisfies some conditions which depend on the ratio
$r\backslash s$.

\begin{prop}\label{prop11}Suppose that $r$, $s$, $t$ are non-zero,
$s=t$, and let $r\backslash s=q_1$. Let $f$ be a homogeneous
Darboux polynomial of degree $m$ which does not have the form
$(\ref{16})$, and let $\alpha _1$, $\alpha _2$, $\beta _1$, $\beta
_2$, $\gamma _1$, $\gamma _2$ be the integers which appear in
propositions $\ref{prop5}$ and $\ref{prop6}$. For
$j=0,1,2,...,m-1$, suppose that $\alpha _1+(\alpha _2-j)\frac
1{q_1}\notin {\cal N}_{m-j}$, $\beta _1-(\beta _2-j)\frac
1{q_1}\notin {\cal N}_{m-j}$, $(\alpha _1-j)q_1+\alpha _2\notin
{\cal N}_{m-j}$, and $-((\beta _1-j)q_1-\beta _2)\notin {\cal
N}_{m-j}$. Then, we have $(i)\hspace{2mm}\alpha _2=\beta _2$ and
$(ii)\hspace{2mm}\alpha _1+\beta _1<\gamma _1+\gamma _2$.
\end{prop}
\hspace{-6mm} Proof. Relation $(i)$ is statement $(a)$ of
proposition \ref{prop7}. We prove the inequality $(ii)$. Suppose
on the contrary that $\alpha _1+\beta _1>\gamma _1+\gamma _2$. By
arguments that we have used repeatedly in this paper (for example
see proposition \ref{prop8}), $f$ is divisible $\alpha _1$ times
by $x_2$ and $\beta _1$ times by $x_1$. Thus we have $f=x_1^{\beta
_1}x_2^{\alpha _1}f^{\prime}$, where $f^{\prime}$ is a Darboux
polynomial of degree $m-(\alpha _1+\beta _1)$ such that\[L
(f^{\prime})=(-\alpha _2sx_1-\beta _2sx_2+(\gamma -(\alpha
_1+\beta _1)s)x_3)f^{\prime}.\]By proposition \ref{prop5} there
exist non-negative integers $\gamma _1^{\prime}$, $\gamma
_2^{\prime}\in\{ 0,1,...,m-(\alpha _1+\beta _1)\}$ such that
$\gamma (f^{\prime})=\gamma _1^{\prime}s+\gamma
_2^{\prime}t=(\gamma _1^{\prime}+\gamma _2^{\prime})s$. This
implies that $\gamma _1^{\prime}+\gamma _2^{\prime}=\gamma
_1+\gamma _2-\alpha _1-\beta _1<0$, a contradiction.

If $\alpha _1+\beta _1=\gamma _1+\gamma _2$, then from the
equation above we have $L(f^{\prime})=(-\alpha _2sx_1-\beta
_2sx_2)f^{\prime}$, and our assumptions imply that $f^{\prime}$ is
divisible $\alpha _2$ times by $x_3$ (proposition \ref{prop8}). So
we have $f^{\prime}=x_3^{\alpha _2}I$, and therefore $f=x_1^{\beta
_1}x_2^{\alpha _1}x_3^{\alpha _2}I$, where $I$ is a first
integral. Since we assume that $f$ does not have the form
(\ref{16}) we may exclude this possibility, and the proof is
completed.\hfill$\Box$

\begin{prop}\label{prop12} Suppose that $r$, $s$, $t$ are non-zero,
 $s=t$, and let $r\backslash s=q_1$. Let $f$ be a homogeneous
Darboux polynomial of degree $m$ which does not have the form
$(\ref{16})$. With the same assumptions as in proposition
$\ref{prop11}$ we have $f=(x_1+x_2)f_1$, for some polynomial
$f_1$.
\end{prop}
\hspace{-6mm}Proof. From proposition \ref{prop8} it follows that
we may assume $\gamma (f)\not= 0$. By proposition \ref{prop3} $f$
does not have an $x_3^m$ term and we can write $f=x_1\phi
_1+x_2\psi _1$, for some polynomials $\phi _1$, $\psi _1$. For a
polynomial $f=f(x_1,x_2,x_3)$ we denote by $\widetilde f$ the
polynomial obtained from $f$ by setting $x_2=-x_1$, that is
$\widetilde f=\widetilde f(x_1,x_3)=f|_{x_2=-x_1}$. So,
$\widetilde f=x_1(\widetilde\phi _1-\widetilde\psi _1)$, and
letting $h_1=\phi _1-\psi _1$ we have $\widetilde f=x_1\widetilde
h_1$. Setting $s=t$ and $x_2=-x_1$ in equation (\ref{8}) we obtain
\[x_1\widetilde {L(\phi _1)}-x_1\widetilde {L(\psi _1)}=((\alpha
-\beta +r)x_1+(\gamma -s)x_3)x_1(\widetilde\phi _1-\widetilde\psi
_1)\]or\begin{equation}\label{18}\widetilde {L(h_1)}=((\alpha
-\beta +r)x_1+(\gamma -s)x_3)\widetilde h_1.\end{equation}Setting
$s=t$ and $x_2=-x_1$ in equation (\ref{5}) we
obtain\begin{equation}\label{19}\widetilde {L
(h_1)}=-x_1(rx_1-sx_3)\Big (\widetilde { \frac {\partial
h_1}{\partial x_1}}-\widetilde {\frac {\partial h_1}{\partial
x_2}}\Big ).\end{equation}Combining equations (\ref{18}) and
(\ref{19}), and noting that $\alpha _2=\beta _2$ (proposition
\ref{prop11}), we
obtain\begin{equation}\label{20}-x_1(rx_1-sx_3)\Big (\frac
{\widetilde {\partial h_1}}{\partial x_1}-\frac {\widetilde
{\partial h_1}}{\partial x_2}\Big )=(-(\alpha _1+\beta
_1-1)rx_1+(\gamma _1+\gamma _2-1)sx_3)\widetilde
h_1.\end{equation} From proposition \ref{prop11} we also have
$\alpha _1+\beta _1<\gamma _1+\gamma _2$, which implies that the
term $-(\alpha _1+\beta _1-1)rx_1+(\gamma _1+\gamma_2-1)sx_3$ is
not a constant multiple of $(rx_1-sx_3)$. Since $(rx_1-sx_3)$
divides the right-hand side of equation (\ref{20}), it divides
$\widetilde h_1$. Therefore we have\[h_1=(rx_1-sx_3)\rho
_1+(-rx_2-sx_3)\chi _1,\]for some polynomials $\rho _1$, $\chi
_1$. Let $h_2=\rho _1+\chi _1$. Then, $\widetilde
h_1=(rx_1-sx_3)\widetilde h_2$ and $\widetilde
f=x_1(rx_1-sx_3)\widetilde h_2$. We
have\begin{equation}\label{21}\frac {\widetilde {\partial
h_1}}{\partial x_1}=(rx_1-sx_3)\frac {\widetilde {\partial\rho
_1}}{\partial x_1}+r\widetilde\rho _1+(rx_1-sx_3)\frac {\widetilde
{\partial\chi _1}}{\partial
x_1},\end{equation}\begin{equation}\label{22}\frac {\widetilde
{\partial h_1}}{\partial x_2}=(rx_1-sx_3)\frac {\widetilde
{\partial\rho _1}}{\partial x_2}+(rx_1-sx_3)\frac {\widetilde
{\partial\chi _1}}{\partial \chi _2}-r\widetilde\chi
_1.\end{equation} Substituting for $\frac {\widetilde {\partial
h_1}}{\partial x_1}$, $\frac {\widetilde {\partial h_1}}{\partial
x_2}$ from equations (\ref{21}), (\ref{22}) respectively in
equation (\ref{20}) we obtain\[-x_1(rx_1-sx_3)((rx_1-sx_3)\Big
(\frac {\widetilde {\partial\rho _1}}{\partial x_1}+\frac
{\widetilde {\partial\chi _1}}{\partial x_1}\Big )-(rx_1-sx_3)\Big
(\frac {\widetilde {\partial\rho _1}}{\partial x_2}+\frac
{\widetilde {\partial\chi _1}}{\partial x_2}\Big
)+\]\[+r(\widetilde\rho _1+\widetilde\chi _1))=(-(\alpha _1+\beta
_1-1)rx_1+(\gamma _1+\gamma _2-1)sx_3)\widetilde h_1\]and
simplifying further we
have\begin{equation}\label{23}-x_1(rx_1-sx_3)\Big (\frac
{\widetilde {\partial h_2}}{\partial x_1}-\frac {\widetilde
{\partial h_2}}{\partial x_2}\Big )=(-(\alpha _1+\beta
_1-2)rx_1+(\gamma _1+\gamma _2-1)sx_3)\widetilde
h_2.\end{equation}The term $-(\alpha _1+\beta _1-2)rx_1+(\gamma
_1+\gamma _2-1)sx_3$ is not a constant multiple of $(rx_1-sx_3)$,
and so $(rx_1-sx_3)|\widetilde h_2$. Continuing in the same way we
find that $\widetilde f$ is divisible by an infinity of powers of
$(rx_1-sx_3)$, which is a contradiction. Therefore we must have
$\widetilde f=0$. This implies that $f=(x_1+x_2)f_1$, for some
polynomial $f_1$, and the proof of the proposition is
completed.\hfill$\Box$

\begin{cor}\label{cor1}Suppose that $r$, $s$, $t$ are non-zero,
 $s=t$, and let $r\backslash s=q_1$. Let $f$ be a homogeneous
Darboux polynomial of degree $m$. With the same assumptions as in
proposition $\ref{prop11}$ we
have\begin{equation}\label{24}f=x_1^ix_2^jx_3^k(x_1+x_2)^lI,
\end{equation}where $I$ is a first integral and $i$, $j$, $k$, $l$
are non-negative integers.
\end{cor}
\hspace{-6mm}Proof. Note first that if $\gamma (f)=0$ then by
proposition \ref{prop8} it follows that we must have $\alpha
_1=\beta _1=0$ and $f=x_3^{\alpha _2}I$. If $\gamma (f)\not= 0$
and $f$ does not have the form (\ref{16}) (which is (\ref{24})
with $l=0$), then by proposition \ref{prop12} we have
$f=(x_1+x_2)f_1$ for some polynomial $f_1$, and \[L (f_1)=(\alpha
x_1+\beta x_2+(\gamma -s)x_3)f_1.\]Repeating this procedure a
finite number of steps, we find that $f$ has the form
(\ref{24}).\hfill$\Box$

\begin{rem}\label{rem4} Similar results hold when $r=s$
and $r=-t$. It can be seen that if $r=s$ then $x_2+x_3$ is a
linear Darboux polynomial with cofactor $-rx_1$. Under conditions
analogous to the ones we have used in this section, we have
$f=x_1^ix_2^jx_3^k(x_2+x_3)^lI$. Similarly, if $r=-t$ then
$x_1+x_3$ is a linear Darboux polynomial with cofactor $rx_2$, and
we have $f=x_1^ix_2^jx_3^k(x_1+x_3)^kI$.
\end{rem}

\section{Examples}

\subsection{Example: The KM-system}

We give a complete description of Darboux polynomials for the case
of the KM system ($r=1$, $s=0$, $t=1$).

\be \label{kmeq}
\begin{array}{lcl}
 \dot x_1&=&x_1x_2  \cr
 \dot x_2&=&-x_1x_2+x_2x_3\cr
 \dot x_3&=&-x_2x_3
\end{array}  \ .
 \ee

The Hamiltonian description of system (\ref{a1}) can be found in
\cite{fadeev} and \cite{damianou1}. We will follow
\cite{damianou1} and use the Lax pair of that reference. The Lax
pair in the case $n=3$ is given by

\begin{equation*}
\dot{L}=[B, L],
\end{equation*}
where

\begin{equation*}
 L= \begin{pmatrix} x_1 & 0 & \sqrt{x_1 x_2} & 0  \cr 0
& x_1 +x_2 & 0& \sqrt{x_2 x_3}   \cr
 \sqrt{x_1 x_2} & 0 & x_2 +x_3 & 0  \cr
 0 & \sqrt{x_2 x_3} & 0& x_3 \end{pmatrix}
\end{equation*}
and

\begin{equation*}
 B=\begin{pmatrix} 0 & 0 & { 1 \over 2} \sqrt{x_1 x_2} & 0 \cr
 0 & 0 & 0&{ 1 \over 2} \sqrt{x_2 x_3}   \cr
 -{ 1\over 2} \sqrt{x_1 x_2} & 0 & 0 & 0  \cr
 0 & -{ 1\over 2} \sqrt{x_2 x_3} & 0&0 \end{pmatrix} \ .
 \end{equation*}
This is an example of an isospectral deformation; the entries of
$L$ vary over time but the eigenvalues  remain constant. It
follows that the  functions $ H_i={1 \over i} {\rm Tr} \, L^i$ are
constants of motion.  We note that \bd H_1= 2 (x_1+x_2+x_3) \ \ed
corresponds to the total momentum and
 \bd H_2= \sum_{i=1}^3 x_i^2+
2 \sum_{i=1}^{2} x_i x_{i+1} \ . \ed

Using (\ref{poisson})  we define the following quadratic Poisson
bracket,  $ \{x_i, x_{i+1} \}=x_i x_{i+1}, i=1, 2, $ and $\{x_1,
x_3 \}=0$. For this bracket det$L=x_1^2 x_3^2$ is a Casimir and
the eigenvalues of $L$ are in involution. Taking the function
$H_1=x_1+x_2+x_3 $ as the Hamiltonian we obtain equations
(\ref{kmeq}). Therefore the system has a Casimir given by $F=x_1
x_3$ and a constant of motion $x_1+x_2+x_3$ corresponding to the
Hamiltonian. Note that $H_2=H_1^2-2 F$.

In the following table we present all Darboux polynomials of
degree $\le 3$ and the corresponding  cofactors.

\begin{table}[h]
\begin{center}
\begin{tabular}{|l|c|c|}
\hline & Darboux polynomial & cofactor \\
\hline
1 & $x_1$ & $x_2$ \\
\hline
2 & $x_2$ & $-x_1+x_3$ \\
\hline
3 & $x_3$ & $-x_2$ \\
\hline 4 & $x_1+x_2+x_3$ & 0\\
\hline
\end{tabular}
\caption {Linear Darboux polynomials and corresponding cofactors}
\end{center}
\end{table}

\begin{table}[h]
\begin{center}
\begin{tabular}{|l|c|c|l|c|c|}
\hline
 & Darboux polynomial & cofactor & & Darboux polynomial & cofactor \\
\hline
1 & $x_1^2$ & $2x_2$ & 6 & $x_2x_3$ & $-x_1-x_2+x_3$\\
\hline
2 & $x_2^2$ & $-2x_1+2x_3$ & 7 & $x_1(x_1+x_2+x_3)$ & $x_2$\\
\hline
3 & $x_3^2$ & $-x_2$ & 8 & $x_2(x_1+x_2+x_3)$ & $-x_1+x_3$\\
\hline
4 & $x_1x_2$ & $-x_1+x_2+x_3$ & 9 & $x_3(x_1+x_2+x_3)$ & $-x_2$ \\

\hline
5 & $x_1x_3$ & 0 & 10 & $c_1(x_1^2+x_2^2+x_3^2+$ & 0\\
 & & & & $+2x_1x_2+2x_2x_3)+c_2x_1x_3$ & \\
\hline
\end{tabular}
\caption{Quadratic Darboux polynomials and corresponding
cofactors. Note that (10) is a sum of two first integrals, and
thus a first integral; $c_1$, $c_2$ are constants}
\end{center}
\end{table}

\begin{table}[h]
\begin{center}
\begin{tabular}{|l|c|c|l|c|c|}
\hline
& Darboux polynomial & cofactor & & Darboux polynomial & cofactor \\
\hline
   1 & $x_1^3$ & $3x_2$ & 9 & $x_2^2(x_1+x_2+x_3)$ & $-2x_1+2x_3$ \\
\hline
  2 & $x_2^3$ & $-3x_1+3x_3$ &  10 & $x_3^2(x_1+x_2+x_3)$ & $-2x_2$\\
\hline
  3 & $x_3^3$  & $-3x_2$ & 11 & $x_1x_2(x_1+x_2+x_3)$ & $-x_1+x_2+x_3$  \\
\hline
  4 & $x_1^2x_2$ & $-x_1+2x_2+x_3$  &  12 & $x_2x_3(x_1+x_2+x_3)$  & $-x_1-x_2+x_3$  \\
\hline
  5 & $x_2^2x_1$  & $-2x_1+x_2+2x_3$  &  13 & $c_1x_1(x_1^2+x_2^2+x_3^2+$  & $x_2$  \\
 & & & & $+2x_1x_2)+c_2x_1^2x_3$ &\\
\hline
 6 & $x_2^2x_3$ & $-2x_1-x_2+2x_3$ & 14 &$c_3x_2(x_1^2+x_2^2+x_3^2+$  & $-x_1+x_3$\\
   &  &  &  & $+2x_1x_2+2x_2x_3)+c_4x_1x_2x_3$ &\\
\hline
 7 & $x_3^2x_2$ & $-x_1-2x_2+x_3$ &  15 & $c_5x_3(x_1^2+x_2^2+x_3^2$ & $-x_2$ \\
   &  &  &  & $+2x_1x_2+2x_2x_3)+c_6x_3^2x_1$ &\\
\hline
 8 & $x_1^2(x_1+x_2+x_3)$ & $2x_2$ & & &\\
\hline
\end{tabular}
\caption{Cubic Darboux polynomials and corresponding cofactors;
$c_1,...,c_6$ are constants}
\end{center}
\end{table}

\subsection{Periodic KM-system}
The periodic system ($r=1$, $s=-1$, $t=1$)

The periodic KM-system is given with the same equations (\ref{a1})
plus a periodicity condition $x_i=x_{i+n}$.  In the case $n=3$ we
obtain:
\[\hspace{8mm}\dot x_1=x_1x_2-x_1x_3\]\[\hspace{11mm}\dot x_2=-x_1x_2+x_2x_3\]\[\hspace{8mm}\dot
x_3=x_1x_3-x_2x_3\vspace{5mm}\]

We give a different type of Lax  pair for this system from
\cite{AVV}.

\bd L= \begin{pmatrix}
  0 & x_1 & 1 \cr
  1 & 0 & x_2 \cr
  x_3 & 1 & 0  \end{pmatrix}
\ed

\bd B=\begin{pmatrix}
  0 & 0 & x_1x_2 \cr
  x_2x_3 & 0 & 0  \cr
  0 & x_3x_1 & 0 \end{pmatrix}
\ . \ed

 It follows that the functions $ H_i={1 \over i} {\rm Tr} \, L^i$
are constants of motion.  We note that $ H_1= 0 \ $, $ H_2=
x_1+x_2+x_3 $ and $H_3=1+ x_1 x_2 x_3$.  As expected the function
$H_2=x_1+x_2+x_3$ plays the role of the Hamiltonian with respect
to the poisson bracket (\ref{poisson}) while $F=x_1 x_2 x_3$ is a
Casimir.

In the following table  we present all Darboux polynomials of
degree $\le 3$ and the corresponding  cofactors.

\begin{table}[h]
\begin{center}
\begin{tabular}{|l|c|c|}
\hline & Darboux polynomial & cofactor \\
\hline
1 & $x_1$ & $x_2-x_3$ \\
\hline
2 & $x_2$ & $-x_1+x_3$ \\
\hline
3 & $x_3$ & $x_1-x_2$ \\
\hline 4 & $x_1+x_2+x_3$ & 0 \\
\hline
\end{tabular}
\caption {Linear Darboux polynomials and corresponding cofactors}
\end{center}
\end{table}

\begin{table}[h]
\begin{center}
\begin{tabular}{|l|c|c|l|c|c|}
\hline
 & Darboux polynomial & cofactor & & Darboux polynomial & cofactor \\
\hline
1 & $x_1^2$ & $2x_2-2x_3$ & 6 & $x_2x_3$ & $-x_2+x_3$\\
\hline
2 & $x_2^2$ & $-2x_1+2x_3$ & 7 & $x_1(x_1+x_2+x_3)$ & $x_2-x_3$\\
\hline
3 & $x_3^2$ & $2x_1-2x_2$ & 8 & $x_2(x_1+x_2+x_3)$ & $-x_1+x_3$\\
\hline
4 & $x_1x_2$ & $-x_1+x_2$ & 9 & $x_3(x_1+x_2+x_3)$ & $x_1-x_2$ \\
\hline
5 & $x_1x_3$ & $x_1-x_3$ & 10 & $(x_1+x_2+x_3)^2$ & 0 \\
\hline
\end{tabular}
\caption{Quadratic Darboux polynomials and corresponding
cofactors}
\end{center}
\end{table}

\begin{table}[h]
\begin{center}
\begin{tabular}{|l|c|c|l|c|c|}
\hline
& Darboux polynomial & cofactor & & Darboux polynomial & cofactor \\
\hline
   1 & $x_1^3$ & $3x_2-3x_3$ & 11 & $x_2(x_1+x_2+x_3)^2$ & $-x_1+x_3$ \\
\hline
  2 & $x_2^3$  & $-3x_1+3x_3$  &  12 & $x_3(x_1+x_2+x_3)^2$ & $x_1-x_2$\\
\hline
  3 & $x_3^3$  & $3x_1-3x_2$ & 13  & $x_1^2(x_1+x_2+x_3)$ & $2x_2-2x_3$  \\
\hline
  4 & $x_1^2x_2$ & $-x_1+2x_2-x_3$  &  14 &  $x_2^2(x_1+x_2+x_3)$ & $-2x_1+2x_3$  \\
\hline
  5 & $x_1^2x_3$  & $x_1+x_2-2x_3$  &  15 & $x_3^2(x_1+x_2+x_3)$  & $2x_1-2x_2$  \\
\hline
 6 & $x_2^2x_1$ & $-2x_1+x_2+x_3$ & 16 & $x_1x_2(x_1+x_2+x_3)$ & $-x_1+x_2$ \\
\hline
 7 & $x_2^2x_3$ & $-x_1-x_2+2x_3$ & 17 & $x_1x_3(x_1+x_2+x_3)$ & $x_1-x_3$\\
\hline
 8 & $x_3^2x_1$ & $2x_1-x_2-x_3$ & 18 & $x_2x_3(x_1+x_2+x_3)$ & $-x_2+x_3$ \\
\hline
 9 & $x_3^2x_2$ & $x_1-2x_2+x_3$ & 19 & $x_1x_2x_3$ & 0 \\
 \hline
 10 & $x_1(x_1+x_2+x_3)^2$ & $x_2-x_3$ & 20 & $c_1(x_1^3+x_2^3+x_3^3+3x_1^2x_2+$ & 0\\
 & & & & $+3x_1^2x_3+3x_2^2x_1+3x_2^2x_3+$ &\\
 & & & & $+3x_3^2x_1+3x_3^2x_2)+c_2x_1x_2x_3$ &\\
\hline
\end{tabular}
\caption{Cubic Darboux polynomials and corresponding cofactors;
$c_1$, $c_2$ are constants}
\end{center}
\end{table}

\subsection{
The case $s=t$ $(r=5,s=t=1)$ }

\[\hspace{8mm}\dot x_1=5x_1x_2+x_1x_3\]\[\hspace{11mm}\dot x_2=-5x_1x_2+x_2x_3\]\[\hspace{11mm}\dot
x_3=-x_1x_3-x_2x_3\vspace{5mm}\]

We list below all linear, quadratic, and cubic Darboux polynomials
of the above system which do not have the form (\ref{16}), and
their corresponding cofactors.\vspace{5mm}

\begin{table}[h]
\begin{center}
\begin{tabular}{|l|c|c|}
\hline & Darboux polynomial & cofactor \\
\hline
1 & $x_1+x_2$ & $x_3$ \\
\hline
\end{tabular}
\caption {Linear Darboux polynomials and corresponding cofactors}
\end{center}
\end{table}

\begin{table}[h]
\begin{center}
\begin{tabular}{|l|c|c|l|c|c|}
\hline
 & Darboux polynomial & cofactor & & Darboux polynomial & cofactor \\
\hline
1 & $x_1(x_1+x_2)$ & $5x_2+2x_3$ & 4 & $(x_1+x_2)^2$ & $2x_3$\\
\hline
2 & $x_2(x_1+x_2)$ & $-5x_1+2x_3$ & 5 & $(x_1+x_2)(x_1+x_2+x_3)$ & $x_3$\\
\hline
3 & $x_3(x_1+x_2)$ & $-x_1-x_2+x_3$ &  & & \\
\hline
\end{tabular}
\caption{Quadratic Darboux polynomials and corresponding
cofactors}
\end{center}
\end{table}

\begin{table}[h]
\begin{center}
\begin{tabular}{|l|c|c|l|c|c|}
\hline
& Darboux pol. & cofactor & & Darboux polynomial & cofactor \\
\hline
   1 & $(x_1+x_2)^3$ & $3x_3$ & 9 & $x_1x_3(x_1+x_2)$ & $-x_1+4x_2+2x_3$ \\
\hline
  2 & $x_1(x_1+x_2)^2$  & $5x_2+3x_3$  &  10 & $x_2x_3(x_1+x_2)$ & $-6x_1-x_2+2x_3$\\
\hline
  3 & $x_2(x_1+x_2)^2$  & $-5x_1+3x_3$ & 11  & $x_1(x_1+x_2)(x_1+x_2+x_3)$ & $5x_2+2x_3$  \\
\hline
  4 & $x_3(x_1+x_2)^2$ & $-x_1-x_2+2x_3$  &  12 & $x_2(x_1+x_2)(x_1+x_2+x_3)$  & $-5x_1+2x_3$  \\
\hline
  5 & $x_1^2(x_1+x_2)$  & $10x_2+3x_3$  &  13 & $x_3(x_1+x_2)(x_1+x_2+x_3)$  & $-x_1-x_2+x_3$  \\
\hline
 6 & $x_2^2(x_1+x_2)$ & $-10x_1+3x_3$ & 14 & $(x_1+x_2)(x_1+x_2+x_3)^2$ & $x_3$ \\
\hline
 7 & $x_3^2(x_1+x_2)$ & $-2x_1-2x_2+x_3$ & 15 & $(x_1+x_2)^2(x_1+x_2+x_3)$ & $2x_3$ \\
\hline
 8 & $x_1x_2(x_1+x_2)$ & $-5x_1+5x_2+3x_3$ &  & & \\
\hline
\end{tabular}
\caption{Cubic Darboux polynomials and corresponding cofactors}
\end{center}
\end{table}

\section{Acknowledgement}
We thank the Cyprus Research Promotion Foundation for support
through the grant CRPF0506/03.

\end{document}